\documentclass[preprint,aps,floatfix,showpacs]{revtex4}
\usepackage{graphicx,amssymb,amsmath}
\usepackage[usenames]{color}

\usepackage{lscape}
\begin{document}
\preprint{\vbox{\hbox{ JLAB-THY-12-1514} }}
\title{\phantom{x}
\vspace{-0.5cm}     }
\title{Holographic models and the QCD trace anomaly}
\author{
 J.~L.~Goity $^{a,b}$ \thanks{e-mail: goity@jlab.org} and
R. C. Trinchero $^{c,d}$ \thanks{e-mail: trincher@cab.cnea.gov.ar }}

\affiliation{$^a$Department of Physics, Hampton University, Hampton, VA 23668, USA.\\
$^b$Thomas Jefferson National Accelerator Facility, Newport News, VA 23606, USA.\\
$^c$Instituto Balseiro, Centro At\'omico Bariloche,
8400 San Carlos de Bariloche, Argentina.\\
$^d$CONICET, Rivadavia 1917,  1033  Buenos Aires, Argentina.}
 
\begin{abstract}
Five dimensional dilaton models are considered as possible holographic duals of the pure gauge QCD vacuum. In the framework of these models, the QCD trace anomaly equation is considered. Each quantity appearing in that equation is computed by holographic means. Two exact solutions for different dilaton potentials corresponding to perturbative and non-perturbative $\beta$-functions are studied. It is shown that in the perturbative case, where the $\beta$-function is the QCD one at leading order, the resulting space is not asymptotically AdS. In the non-perturbative case, the model considered presents confinement of static quarks and leads to a non-vanishing gluon condensate, although it does not correspond to an asymptotically free theory. Calculating the Nambu-Goto action corresponding to a small circular Wilson loop, leads to an expression for the gluon condensate. The validity of the trace anomaly equation is considered for both models. It holds for the perturbative model and it does not hold for the non-perturbative one.
\end{abstract}


\maketitle

\section{Introduction 
}
The relation between large N  gauge theories and string theory \cite{tHooft} together with the anti-de Sitter/conformal field theory (AdS/CFT) correspondence \cite{Malda, Gubser, Witten, Malda-Review} have opened new insights into strongly interacting gauge theories. The application of these ideas to QCD has received significant attention since those breakthroughs. From the  phenomenological point of view, the so called AdS/QCD approach has  produced very interesting results in spite of the strong assumptions involved in its formulation \cite{AdS/QCD}. It seems important to further proceed investigating these ideas and refining the current understanding of a possible QCD gravity dual.  This endeavor has been followed in references \cite{Gursoy-Kiritsis}. The aim of the present paper is to explore the simplest non-perturbative features of QCD. This is done in the framework of a holographic description of the pure Yang-Mills (YM) QCD vacuum by means of  5-dimensional dilaton gravity models.

At the basis of the AdS/CFT correspondence is the connection between scale transformations in the boundary field theory and isometries of the bulk gravitational theory.  However, QCD is not a conformal field theory,  as  the scale symmetry is broken by the trace  anomaly \cite{TraceAnomaly}. The trace anomaly equation  describes the behavior of QCD under scale transformations.  The question to be explored is how   a holographic model can incorporate this behavior.  

The trace anomaly equation  \cite{TraceAnomaly} states that,
\begin{equation}\label{trace-anomaly}
T_{i}^{i}=\frac{\beta(\lambda)}{\lambda} {\rm Tr} (G_{ij}G^{ij}) 
\end{equation}
 where $T_{i}^{i}$
  denotes the trace of the QCD energy momentum tensor (latin indices for space-time), $\beta(\lambda)$
  is the QCD $\beta$-function,  $\lambda=N\,\frac{g_{YM}^{2}}{4\pi}$
  is the t'Hooft coupling, $G_{ij}$
  is the QCD field strength tensor and the trace is taken in the fundamental representation of  the SU(N)
  gauge group. In this respect it is important to note that holographic models can tell something about each of the three quantities involved in the trace anomaly equation, namely the vacuum expectation value (VEV)    of the trace of the energy momentum tensor,   the $\beta$-function and  the VEV of ${\rm Tr}(G_{ij}G^{ij})$.  
  
 According to the correspondence, evaluating the five dimensional action at a classical global solution gives information about the  VEV  of the trace of the energy momentum tensor. The $\beta$-function can be obtained in terms of the solutions to the 5-dimensional equation of motion derived form the action in the bulk.
 Finally, there is a way of calculating the VEV  of the Wilson loop by means of minimizing the Nambu-Goto (NG) action for a loop lying in the boundary space. This is known to work in the strictly AdS case, i.e. for a conformal boundary field theory, and its generalization to non-conformal cases  is  still an open important problem. In turn, the VEV  
defined by $G_2\equiv \frac{g_{YM}^2}{4\pi^2}\langle G_{ij}G^{ij}\rangle$, known as the gluon condensate, 
 can be determined  from  the  coefficient of the area squared in the expansion of a small Wilson loop in powers of its area \cite{Banks-et-al,DiGiacomo-et-al,Rakow}.

The  features and results of this work are summarized as follows,
 \begin{itemize}
\item Two exact solutions of 5-dimensional dilaton gravity for different dilaton   potentials are considered. The first model, to  be referred to as perturbative model, has a $\beta$-function, which to leading order in the t'Hooft coupling is the same as the perturbative 1-loop QCD $\beta$-function.
The second model  will be referred to as non-perturbative model (because its $\beta$-function is non-analytic in $\lambda$). This model, by choice of the parameter $\alpha$ in the model,  can be made to correspond    asymptotically to the soft wall model often used in non-dynamical  models of  holographic QCD.  The model leads naturally to confinement in the sense of static quarks, and to a non-vanishing gluon condensate when tested with a Wilson loop. However,  it does not lead to asymptotic freedom in the ultraviolet.

\item  For the perturbative model the asymptotic behavior of the solutions in the ultraviolet is not AdS. In the language of the holographic renormalization group the difference with the AdS limit is produced by an irrelevant operator that flows away from the AdS fixed point. In the  non-perturbative model considered,  the $\beta$-function  gives rise to an UV fixed point at finite $\lambda$ and the metric is asymptotically AdS.

\item Using the correspondence, the VEV of the energy momentum tensor is obtained by evaluating the 5-dimensional action in the corresponding exact solutions, regularizing by introducing an  energy scale and subtracting. These subtractions are performed  as proposed in \cite{haw-horo96}, and employed in the holographic case in \cite{myers99}. In the perturbative case,  taking into account Eqn. (\ref{trace-anomaly}),  it is argued that  the same solution should be subtracted, leading to a vanishing VEV for the energy-momentum tensor. In the non-perturbative model, being asymptotically AdS,  the AdS limit is subtracted.

\item In order to calculate the gluon condensate  the VEV of a small circular Wilson loop is considered. This is carried out  using the corresponding  NG action. For the perturbative model this procedure leads to a vanishing gluon condensate, while a non-vanishing result is obtained in the non-perturbative case.

\item The validity of  Eqn.(\ref{trace-anomaly}) is considered for both models, and shown to hold in the perturbative one. 
 In the non-perturbative model the dependence of the gluon condensate on the energy scale is not the one required by  Eqn. (\ref{trace-anomaly}). This is however not  unexpected  as  this model does not give a consistent description of the QCD ultraviolet behavior.
\end{itemize}

The paper is organized as follows. Section II presents the 5-dimensional dilaton-gravity model employed in what follows. Exact solutions of the dilaton model  equations of motion and associated $\beta$-functions corresponding to the perturbative and non-perturbative models are studied in section III. Section IV deals with the evaluation, regularization and subtraction of the gravitational action evaluated in the above mentioned exact solutions. Section V discusses the relevant asymptotics of the  solutions of section III, and gives the explicit result for the subtracted gravitational action for those solutions. Section VI presents a study of the VEV of a small circular Wilson loop by means of the minimization of the NG action. Section VII addresses the issue of validity of the trace anomaly equation in the models considered.    A final section VIII presents   conclusions and outlook.

\section{Dilaton Model}
 The model   considered is that of a self interacting scalar field immersed in a dynamical gravitational field in $d+1$
  dimensions (in the end the results are only valid at $d=4$). The action of the model is given by \cite{DilatonAction},
\begin{equation}
S_{d+1}=\frac{1}{16\pi\, G_{N}^{(d+1)}}\left(\int_{M_{d+1}}\, d^{d+1}x\;\sqrt{g}\,(-R+\frac{1}{2}\,g^{\mu\nu}\partial_{\mu}\phi\,\partial_{\nu}\phi-V(\phi))-2\int_{M_{d}}d^{d}x\,\sqrt{h}\, K\right),
\end{equation}
 where $G_{N}^{(d+1)}$
  is the  Newton constant  in $d+1$-dimensions (of dimension $(d-1)$), $g_{\mu\nu}$
    the metric tensor field, $R$
    the scalar curvature, $\phi$
   the dilaton field,  and $V(\phi)$
  the dilaton potential. The last term is the Gibbons-Hawking term \cite{Gibbons-Hawking} where $K$
  is the second fundamental form. This term is included  to make the Lagrangian depend only on the first derivatives of the metric.  The equations of motion derived from this action are,
  \begin{eqnarray}
  E_{\mu\nu}-\frac{1}{2}\,\partial_{\mu}\phi\,\partial_{\nu}\phi+\frac{1}{4}g_{\mu\nu}\,(\partial\phi)^{2}-\frac{1}{2}\,g_{\mu\nu}V(\phi)	&=&	0\nonumber\\
\partial_{\mu}(\sqrt{g}\,g^{\mu\nu}\partial_{\nu}\phi)+\sqrt{g}\;\frac{\partial V(\phi)}{\partial\phi}	&=&	0
\end{eqnarray}
 where the Einstein tensor $E_{\mu\nu}$
  reads: $E_{\mu\nu}	=	R_{\mu\nu}-\frac{1}{2}g_{\mu\nu}R$,  and 
$(\partial\phi)^{2}	=	g^{\mu\nu}\partial_{\mu}\phi\partial_{\nu}\phi$.
 Because here the focus is on the vacuum of the boundary field theory, only metrics and scalar fields having flat boundary space isometry invariance are considered, thus only solutions for the metric and scalar field of the following general form are considered,
 \begin{eqnarray}
 ds^{2}	=	 du^{2}+e^{2A(u)}\,\eta_{ij}\,dx^{i}dx^{j},&~~~&\phi	=	\phi(u),
\end{eqnarray}
where $\eta_{ij}$ is a flat metric, and the coordinates employed here are known as domain wall coordinates. The boundary of the space is at $u=\pm\infty$.
  The AdS metric corresponds to taking $A_{AdS}(u)=u$, where the coordinate $u$
  is measured in units of the AdS radius $L$. For this particular choice of fields which only depend on $u$, the equations of motion are given by,
  \begin{eqnarray}\label{motionglobal}
  A''+d\,A'^2-\frac{V(\phi)}{d-1}	&=&	0\nonumber\\
d\,A'^{2}-\frac{\phi'^{2}}{2(d-1)}-\frac{V(\phi)}{d-1}	&=&	0\nonumber\\
\phi''+d\,A'\phi'+\frac{dV(\phi)}{d\phi}	&=&	0,
\end{eqnarray}
 where the prime
  denotes derivation with respect to $u$.  Introducing a superpotential $ W(\phi)$ according to:
\begin{eqnarray}\label{superpotential1}
A'(u)	&=&	W(\phi)\\\label{superpotential2}
\phi'(u)	&=	&\xi\frac{dW(\phi)}{d\phi},
\end{eqnarray}
the choice $\xi=2(1-d)<0$ reduces the three equations in Eqn.  (5) to the single equation:
 \begin{equation}
 \xi\left(\frac{dW(\phi)}{d\phi}\right)^{2}+dW^{2}-\frac{V(\phi)}{d-1}	=	0.
 \end{equation}
  Since the intended realistic application to QCD is at $d=4$, throughout   $\xi=-6$ could be replaced.
  
  \section{  $\beta$-functions in   dilaton models}
   
  In the AdS/CFT correspondence the identification is made of the YM coupling with the dilaton profile according to \cite{Malda-Review}:
  \begin{equation}
\frac{ \lambda}{N}= \frac{g_{YM}^{2}}{4\pi}=e^{\phi}
  \end{equation}
   The energy scale $\mu$  
 (measured in units of a scale  $\frac{1}{L}$, where $L$ is the length unit mentioned earlier) of the boundary theory is identified with the scale  factor  $e^{A(u)}$
  in domain wall coordinates: $\mu=e^{A(u)}$.
 These identifications give the  $\beta$-function in the dilaton model \cite{Gursoy-Kiritsis}:
 \begin{equation}\label{beta}
 \beta(\lambda)=\frac{d\lambda}{d\log\mu} =N e^{\phi}\frac{\phi'}{A'}=\xi\,\lambda\,\frac{\partial}{\partial\phi}\log W(\phi).
 \end{equation}

In the rest of this section two different and  exactly soluble dilaton models are considered. These models are obtained according to the following scheme: a dilaton profile $\phi(u)$
  is given, where by expressing  $\phi'(u)$
   in terms of  $\phi(u)$ and employing   Eqn. (\ref{superpotential2})   the superpotential $W(\phi)$
  is obtained, followed by  integrating  Eqn. (\ref{superpotential1}) to obtain    $A(u)$, and finally  from  Eqn. (\ref{beta})   the $\beta$-function is obtained. The potential  $V(\phi)$ is determined  from  Eqn. (\ref{motionglobal}).
 
The two models considered are extreme cases. One  model corresponds at leading order in the gauge coupling to the perturbative QCD $\beta$-function, while the other one corresponds to a non-perturbative $\beta$-function, i.e., which is non-analytic at small coupling and which leads to an UV fix point. These models  are  qualitatively different as the next sections show. The precise choice of dilaton profiles is made so as to be able to perform all the calculations analytically.

\subsection{ Perturbative   $\beta$-function}

The following dilaton profile is considered,
\begin{equation}
\phi(u)=-\frac{1}{2}\,\log((\alpha\,u)^{2}+\kappa^{2})
\end{equation}
 Note that this choice means that $\lambda\leq N/\kappa$. Therefore, $\kappa$ should be a quantity order $N$.
Using the procedure just described  leads to:
\begin{equation}{\label{pqcd}}
A(u)=A_{0}+A_{1}u-\frac{1}{2\,\xi}\log\left((\alpha\,u)^{2}+\kappa^{2}\right)+\frac{\alpha\,u}{2\,\kappa\,\xi}\arctan\left(\frac{\alpha\,u}{\kappa}\right),
\end{equation}
where for convenience the integration constants can be chosen in such a way that the leading asymptotic behavior be AdS, namely $A_0=\frac{1}{2\,\xi}$, and   $A_1=1- \frac{\alpha\,\pi}{4\,\kappa\,\xi}$. Then, asymptotically  $A(u)\sim   u-\frac{1}{\xi} \log(\alpha\, u) +{\mathcal{O}}(1/u^2)$.
The resulting $\beta$ function reads:

\begin{equation}
\beta(\lambda)=\frac{2   \, \kappa \, \xi\, \lambda
   ^2  \sqrt{N^2-\kappa ^2
   \lambda ^2}}{  \, \kappa \,
   \lambda  \sqrt{N^2-\kappa ^2
   \lambda ^2}+N^2 ( 
   \arcsin\!\left(\frac{\kappa 
   \lambda }{N}\right)-2  \,
  \frac{ \kappa \, \xi }{\alpha})},
   \end{equation}
   which to leading order in $\lambda$ 
   becomes:
  \begin{equation}
  \beta(\lambda)=-\frac{\alpha\lambda^2}{N }+{\mathcal{O}}(\lambda^3).
  \end{equation}
The choice  $\alpha=\frac{11 N}{6\pi}$ reproduces the leading order term of the QCD $\beta$-function (see Fig. 1)
 

\subsection{ Non-perturbative $\beta$-function}

A $\beta$-function with  non-perturbative behavior, i.e. non-analytic in the coupling $\lambda$, is obtained from the following  dilaton profile,
\begin{equation}
\phi(u)=C\, e^{-\alpha\;u}
\end{equation}
where $\alpha>0$. In this case,
\begin{equation}
 A(u)= u+\frac{C^{2}}{4\xi}\, e^{-2\alpha\,u},
 \end{equation}
giving an asymptotically AdS metric.
 
The resulting $\beta$-function is then given by,
\begin{equation}\label{betaNP}
\beta(\lambda)	=	 -	\frac{\alpha\, \lambda \log\frac{\lambda}{N}}{1-\frac{\alpha}{2\xi}\log^2\frac{\lambda}{N}},
\end{equation}
which is positive in the interval $0<\lambda<N$, leading to an  UV fixed point at $\lambda=N$ (see Fig. 1). Thus, this theory is not asymptotically free, and therefore is not related to  a pure YM theory. The sign of the constant $C$ determines two phases of the theory: for $C<0$ the theory becomes free in the infrared, while for $C>0$ it becomes strongly coupled. Indeed this latter case describes a confining theory in the IR.
 In order to see this it is convenient to express the above result in the conformal coordinate $z$, where asymptotically 
 $ u=-\log(z)$,
 Êand therefore  $A(z)=-\log(z)+\frac{C^{2}}{4\xi}z^{2\alpha}$. This $A$ matches the G\"ursoy-Kiritsis \cite{Gursoy-Kiritsis} criterion for confinement \footnote{For  $\alpha=1$, $ A(z)$
  differs from the one considered in \cite{Andreev-Zakharov} only by the sign of the  $z^{2}$
  term. The AZ model is not a dynamical one, in particular in that reference this same factor $ A(z)$
  is employed in the calculation of the Nambu-Goto action, there is no string frame correction due to the dilaton and so the G\"ursoy-Kiritsis confinement criterium can not be applied to that model.}
 The negative sign  of the coefficient  multiplying the $z^{2\alpha }$ term is crucial in two respects:  it is necessary for  the confinement criterion \cite{Gursoy-Kiritsis} to be fulfilled and second, the behavior of the factor $e^{A(z)}$
  for $z\to\infty$
  is such that, $\lim_{z\to\infty}e^{A(z)}=0$, which as shown in the next section, makes the use of a infrared cut-off unnecessary in the evaluation of the 5-dimensional action for this solution.
 \vspace*{1cm}

 
 
\begin{figure}
\centerline{\includegraphics[height=2in]{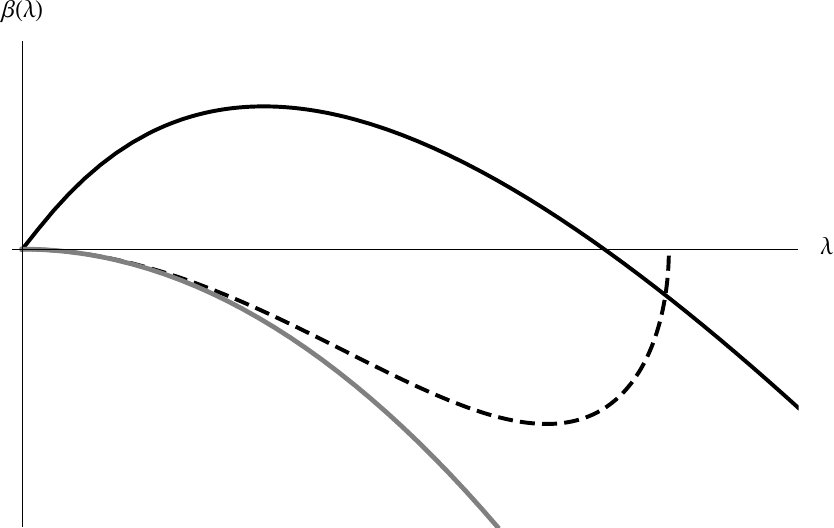}}
\caption{$\beta$ functions of the models considered:  perturbative (dashed), non-perturbative (black), perturbative QCD (gray).  $\beta\equiv 0$ corresponds to the AdS case.}
\end{figure}

\section{The trace of the energy-momentum tensor}

According to the AdS/CFT conjecture, taking the  metric as the source field of the energy-momentum tensor of the boundary field theory,   the VEV of the trace of the energy momentum tensor is evaluated by simply evaluating the action  Eqn. (2) for the classical solutions of the previous section.
 
Taking the trace in the first Eqn. (3)  gives,
\begin{equation}
R=\frac{(d+1)}{(1-d)}\, V(\phi)+\frac{1}{2}(\partial\phi)^{2},
\end{equation}
 and the action for the classical solutions becomes:  
 \begin{equation} S=S_{bulk}+S_{GH}=\frac{1}{16\pi\, G_{N}^{(d+1)}}\int_{M_{d+1}}\, d^{d+1}x\;\sqrt{g}\,\frac{2}{(d-1)}\, V(\phi)+S_{GH},
 \end{equation}
and using the first Eqn. (5),
\begin{equation}
 S_{bulk}=\frac{1}{16\pi \,G_{N}^{(d+1)}}\int_{M_{d+1}}\, d^{d+1}x\;\sqrt{g}\,2(A''+dA'^{2}).
 \end{equation}
 Noting that  $\frac{d^{2}}{du^{2}}e^{dA(u)}=d\;e^{dA(u)}(A''+dA'^{2})$ and $\sqrt{g}=e^{d\,A(u)}$
 leads to,
 \begin{eqnarray}
 S_{bulk}	
	&=&	\frac{V_{M_{d}}}{16\pi\, G_{N}^{(d+1)}}\frac{2}{d}\int_{-\infty}^{\infty}\, du\;\frac{d^{2}}{du^{2}}e^{dA(u)}\nonumber\\
	&=&	\frac{V_{M_{d}}}{8\pi \,G_{N}^{(d+1)}}\frac{1}{d}\left[\frac{d}{du}e^{dA(u)}\right]_{boundary},
	\end{eqnarray}
 where $V_{M_{d}}$ 
   denotes the volume of the boundary d-dimensional space. On the other hand, the classical Gibbons-Hawking boundary action is  given by:
\begin{equation}
 S_{GH}=-\frac{1}{8\pi \, G_{N}^{(d+1)}}\frac{\partial}{\partial n}\int_{M_{d}}d^{d}x\,\sqrt{h},
 \end{equation}
 where $h$
  is the induced metric in the boundary $M_{d}$, namely $\sqrt{h}=e^{dA(u)}$,
  and  $\frac{\partial}{\partial n}$
  denotes a unit vector field orthogonal to the boundary of $M_{d+1}$. In domain wall coordinates this vector field is simply $\frac{\partial}{\partial n}=\frac{\partial}{\partial u}$, and therefore:
 \begin{equation}
 S_{GH}=-\frac{1}{8\pi\, G_{N}^{(d+1)}}V_{M_{d}}\left[\frac{d}{du}e^{dA(u)}\right]_{boundary},
 \end{equation}
 which is just $-d$ times the bulk action.
 For both exact solutions considered in the previous section there is no contribution  from the infrared boundary. 
 On the other hand,  the ultraviolet boundary $u\to\infty$ gives for both cases
   divergent contributions, as it happens in general for any holographic model.
 As proposed in   \cite{Witten},  these contributions can be regularized by evaluating at a finite value $u_{0}$. This leads finally to,
\begin{equation}
S	=	\frac{1}{8\pi\, G_{N}^{(d+1)}}V_{M_{d}}(1-d) e^{dA(u_{0})}A'(u_{0}) .
\end{equation}
 It is important to note that for a boundary theory that is not quantum conformal invariant, as for example QCD, the regulator $u_{0}$
  has a physical meaning. Indeed, as mentioned in the previous section, the energy scale at which the boundary theory is observed is related    to   $u_0$,   the boundary value of the domain wall coordinate $u$.  
  
As shown in  \cite{haw-horo96} and applied to holographic models in \cite{myers99}, a well defined action can be obtained by subtracting from the regulated  action an action corresponding to   some background metric having the same asymptotic limit. 
That is,
\begin{equation}
S_{sub.}=S-S^{asymp.},
\end{equation}
 where $S^{asymp.}$
  denotes the action evaluated in a solution having the same asymptotic behavior as the classical one.
  The subtracted energy-momentum tensor is obtained recalling that, according to the correspondence,
  \begin{equation}
  S=\int_{M_{d}}d^dx\;\sqrt{h}\, h_{ij}T^{ij},
  \end{equation}
 leading to  $T_{i}^{i}(sub.)=\frac{e^{-dA(u_{0})}}{V_{M_{d}}}S_{sub.}$,
 where $A(u_{0})$
  denotes the common asymptotic exponent.
    The choice of this background metric for the solutions considered in section III is discussed in the next section.

\section{ The UV QCD fixed point}

The perturbative model  in  subsection IIIA presents features the understanding of which leads to new insights. These are the following:
\begin{itemize}
\item The model leads to a $\beta$-function that coincides at leading order with the perturbative QCD $\beta$-function.
\item The model is not asymptotically AdS. As Eqn. (\ref{pqcd}) shows, the deviation of $A(u)$ from the AdS limit becomes $\frac{-1}{\xi} \log u$.  
\item As shown in the previous section, the action should be subtracted with the action evaluated in a background metric having the same asymptotic behavior as the one to be subtracted. Thus,  it is not sufficient to perform a substruction with the AdS metric.
\item In the language of the holographic renormalization group \cite{RGE}, this correction corresponds to an irrelevant operator, that flows away from the AdS fixed point \cite{Malda-Review}. This can be seen from the fact that the dilaton field behaves as $-\log u$ at the UV boundary.
 \item   Eqn. (\ref{trace-anomaly}) implies that for QCD the trace of the energy-momentum tensor should vanish in the UV. This can be independently seen in two ways. As shown in section VI for this model,  the VEV of the Wilson loop, calculated via the NG action,  does not have terms which are powers of its area, and therefore  the gluon condensate 
$G_2 $ must vanish. The other way is simply to recall that in perturbative QCD the log of the VEV of the Wilson loop follows a perimeter law.  
\end{itemize}

All these points indicate that the UV fixed point of  QCD does not correspond to AdS. It corresponds to another solution that is well approximated by the one in subsection IIIA  in the UV, i.e. for large $u$, and therefore    the action evaluated in the same solution or one asymptotically equivalent must be subtracted, leading to a vanishing trace of the subtracted energy-momentum tensor.
 
In the non-perturbative model  the space  is asymptotically AdS, and the subtracted action becomes:


 \begin{equation}
S_{sub.}^{NP}=S^{NP}-S^{AdS}=\frac{(1-d)V_{M_{d}}}{8\pi\, G_{N}^{(d+1)}}e^{d\, u_{0}}\left(e^{\frac{d\, C^{2}}{4\,\xi}\, e^{-2\,\alpha\, u_{0}}}(1-\frac{\alpha C^{2}}{2\xi}e^{-2\,\alpha\, u_{0}})-1\right)\;\;, \label{tii-np}
\end{equation}
leading to, 
\begin{equation}
T_{i}^{i}(sub.,NP)=\frac{(1-d)}{8\pi\, G_{N}^{(d+1)}}
\left(1-\frac{\alpha C^{2}}{2\xi}e^{-2\alpha\, u_{0}}-e^{-\frac{d\, C^{2}}{4\,\xi}\, e^{-2\,\alpha\, u_{0}}}\right)\label{Tsub}
\end{equation}

\section{Wilson loops}

The VEV of the operator $G_2$ (gluon condensate) appearing in the trace anomaly is accessible through the power like behavior  of small Wilson loops as a function of their size.  In pure YM theory the expansion of a small smooth Wilson loop (e.g., square or circular) is expected to have the form given by \cite{Banks-et-al, DiGiacomo-et-al,Andreev-Zakharov,Rakow}:
\begin{equation}
\log\langle W(\Gamma)\rangle=-\sum_n C_n9\ell) \left(\frac{\lambda}{N}\right)^n-\frac{\pi^2\, Z}{12\,N} \,G_2\; s^2+\cdots
\end{equation}
where $\ell$ is the length of the loop, $s$ is its area, and   $Z=\beta_1(\lambda)/\beta(\lambda)$ where $\beta_1$  is  the one loop $\beta$-function.  It is argued in pure YM that the terms order $s$ vanish as these would require a gauge invariant dimension two condensate.

The connection between Wilson loops of the boundary conformal gauge theory and minimal surfaces was made in references \cite{Malda-WL,Rey-Yee}. According to it, in a CFT such as ${\mathcal{N}}=4$ SUSY YM, in the large $N$ limit and large 'tHooft coupling the VEV of the Wilson loop is determined by the minimal area surface in the $d+1$ AdS space subtended by the path of the loop $\Gamma$. Specifically:
\begin{eqnarray}
W(\Gamma)&=&\frac{1}{N}\;{\rm Tr\;PExp}(-\oint_\Gamma A_i dx^i)\nonumber\\
\langle W(\Gamma)\rangle &\propto& e^{-S_\Gamma},
\end{eqnarray}
where the minimal area $S_\Gamma$ is given by the  NG action of a string whose ends run along the loop.  Since for a loop located at the boundary  $S_\Gamma$ diverges, it has to be regulated, and thus the proportionality factor above.

The extension of this identification to non-conformal YM theory is  still an open problem, in particular because in that case, as discussed earlier,    the theory cannot be obtained via a relevant  deformation of a CFT \cite{Malda-Review}. This problem is closely related to the problem of finding the non-critical string action for QCD \cite{Polyakov-book}.  An extension of the correspondence for Wilson loops to the non-conformal case has been proposed  \cite{Gursoy-Kiritsis},  in which  the NG action is the one corresponding to the   string-frame metric, namely: $A_S(z)=A(z)+\phi(z)/\sqrt{3}~~$ in $d=4$ dimensions.

For the present purpose  a circular Wilson loop of radius $a$ is considered, for which the NG action turns out to be:
\begin{equation}
S_{NG}=\frac{a^2}{2\pi\, \alpha'}\int_0^1d\rho\;\rho\;e^{2A_S(a\,\omega(\rho))}\;\sqrt{1+\omega'(\rho)^2},
\end{equation}
where $r=a\,\rho$ is the radial coordinate of the disk, and  $z=a\;\omega$ is the bulk coordinate in conformal coordinates. The equation of motion is:
\begin{equation}
\rho\, \omega''+(1+\omega'^2)(\omega'-2\,a\,\rho \,A_S'(a\,\omega))=0,
\end{equation}
where the solution needed satisfies $\omega(1)=0$. In AdS limit it is $\omega(\rho)=\sqrt{1-\rho^2}$, a half sphere.

The UV divergencies of the NG action result from the contributions to the integral for $\rho\to 1$.   Noticing that $\omega'(\rho)$ diverges as $\rho\to 1$, one obtains:
\begin{equation}
\frac{\partial S_{NG}}{\partial z_0}=-\frac{a\, e^{2A_S(z_0)} }{2\pi\,\alpha},
\end{equation}
 where $z_0$ can be interpreted as the location of the loop   in the bulk coordinate $z$ (provided $z_0<<a$).  In  dilaton models one readily obtains: 
\begin{equation}
\frac{\partial S_{NG}}{\partial A_0}= \frac{a\,e^{2A_S(z_0)-A_0}}{2\pi\,\alpha'\,W(A_0)},
\end{equation}
where $A_0\equiv A(z_0)$, which asymptotically for the models discussed $A(z_0)\to -\log z_0$. If the $\beta$-function is given as input to the model,  the superpotential and $A(z)$  are  given by
  \begin{eqnarray}
 W(\lambda)&=&\exp\left(   \frac{1}{2\xi}\int\frac{\beta(\lambda)}{\lambda^2}d\lambda  \right)    \nonumber\\
 A(\lambda)&=&\int\frac{d\lambda}{\beta(\lambda)}  .
 \end{eqnarray}
 One readily checks the AdS case where $W=const$ and $A_S=A$, giving   $ \frac{\partial S_{NG}}{\partial A_0}=\frac{a}{2\pi \,\alpha'}\,e^{ A_0}$.
 
  The perturbative model  asymptotically gives 
   $\beta(\lambda)=-\frac{\alpha}{N}\,\lambda^2$, $\phi(A)=-\log(\alpha A)$ and $W(A)=1-\frac{1}{\xi A} $, where,  without loss of generality,  the constant of integration  required for $W(A)$ has been chosen to be $W_0=1$. This leads to:
 \begin{equation}
  \frac{\partial S_{NG}}{\partial A_0}=\frac{a}{2\pi \,\alpha'}\,\exp\left(A_0-\frac{2}{\sqrt{3}}\log(\alpha A_0)\right).
 \end{equation} 
 This shows that, as one would expect from the fact that the metric is not asymptotically AdS, the UV divergence of the action is modified with respect to the AdS case by the second term in the exponent.

   The non-perturbative model is asymptotically AdS and thus the expectation is that   the UV divergence coincides with the AdS case.
   If the coefficient $\alpha>1$ this is indeed the case as it is easily shown using Eqns. (15) to (17) for $\alpha>1$, which leads to:
   \begin{equation}
    \frac{\partial S_{NG}}{\partial A_0}=\frac{a e^{A_0^{AdS}}}{2\pi \alpha'}\;(1+\frac{2\,C}{\sqrt{3}}e^{-\alpha A_0^{AdS}}+{\cal{O}}(e^{-2\alpha A_0^{AdS}})).
   \end{equation}
   For $\alpha=1$ a constant term remains, which corresponds to a term linear in $A_0$ in the UV divergence of  $S_{NG}$ or equivalently logarithmic in $z_0$.
   
   The UV divergencies stem from the fact that $A_S$ diverges at the boundary. Therefore, they must naturally be only proportional to the perimeter of the loop, i.e.
 proportional to $a$. 
   For this reason, the  contributions  of  higher powers of $a$,  which are of interest here,   are  independent of the regularization of  $S_{NG}$  and  unambiguous.

The central  point  of the discussion  is the sufficient conditions for the presence of higher power terms in $a$ in $S_{NG}$.  The simplest case is when the metric is asymptotically AdS and the UV divergence of $S_{NG}$ corresponds as well to the AdS case. For small $a$,  $A_S(z)=A_{AdS}(z)+\delta\!A(z)$, and expanding in $\delta\!A$ leads to:
 \begin{equation}
S_{NG}=\frac{1}{2\pi\,\alpha'}\int_0^1\frac{d\eta}{\eta^2}\;(1+2\;\delta \!A(a\eta))+{\mathcal{O}}(\delta \!A^2),
\end{equation}
where $\eta=\sqrt{1-\rho^2}$, and evaluation in the AdS limit solution has been performed. 
The first order approximation is adequate near the boundary $\eta\to 0$ only if the UV divergencies are strictly AdS.   On the other hand,  the dependencies of $S_{NG}$ in powers of $a$ beyond the first power (perimeter terms) will stem  primarily from the interior of the integration domain, where the approximation is expected to work. Thus, 
 a sufficient condition for such power corrections is  that $\delta \,A$ contains terms which have power dependency  in the argument. The  contributions ${\mathcal{O}}(\delta \!A^2)
$ in Eqn. (38) are in general difficult to evaluate as they involve the corrections to the solution of the equation of motion (32)  \cite{Andreev-Zakharov}.   The arguments made here apply in particular to the non-perturbative model when $\alpha>1$.                                                                                        
 
When the metric is not asymptotically  AdS, as is the case of the perturbative model,  a more accurate evaluation is necessary.      For $a$ sufficiently small  the entire surface will lye near the boundary $u\to\infty$, and $\kappa$ can be set to zero, thus $\phi(u)=-\log(\alpha u)$, $A(u)=u- \frac{1}{\xi} \log(\alpha u)$. Setting $u=-\log z$ and evaluating $S_{NG}$  with the asymptotic AdS solution $z(\eta)=a \eta$ leads to:
\begin{equation}
S_{NG}=\frac{1}{2\pi\alpha'}\int_0^1\frac{d\eta}{\eta^2} \exp(-(\frac{1}{\xi}+\frac{2}{\sqrt{3}})\log(-\alpha\log(a\eta))).
\end{equation}
It is readily checked that this has the UV divergence obtained earlier in Eqn. (36).
 Evidently the dependence of $S_{NG}$ in $a$ is logarithmic, and therefore according to the evaluation of the Wilson loop $G_2=0$ in the perturbative model.
A similar conclusion results if  $\beta(\lambda)$ is in general analytic in $\lambda$. 
Therefore,  in the present framework, this indicates that in order to obtain a non-vanishing gluon condensate, the $\beta$ function should include non-analytic terms in $\lambda$.

As an illustration of  the latter, where power corrections are obtained at small coupling as consequence of non-perturbative terms in the $\beta$-function, consider the   asymptotically free theory with  $\beta(\lambda)=-b_0\,\lambda^2\,(1+c\,\exp(-\frac{\alpha}{\lambda}))$, which is found in certain SUSY gauge theories \cite{Seiberg} as the result of instanton contributions. Considering the non-perturbative piece as small (or expanding in $c$), asymptotically $W(A)=e^{-\frac{1}{\xi A}}(1-\frac{c}{\xi \alpha b_0 A^2} e^{-\alpha b_0 A})$, $\phi(A)=-\log b_0 A+\frac{c}{\alpha b_0 A} e^{-\alpha b_0 A}$,  leading to:
 \begin{equation}
 \frac{\partial S_{NG}}{\partial A_0}=
 \frac{\partial S_{NG}}{\partial A_0}^{\rm pert.}(A_0) (1+\frac{2c\, e^{-b_0\alpha A_0}}{\alpha b_0 A_0}(\frac{1}{\xi A_0}+\frac{1}{\sqrt{3}})),
  \end{equation}
   which as expected coincides asymptotically with the perturbative model.  The evaluation of the finite pieces gives power terms in $a$.  Asymptotically, to first order in $c$:
\begin{eqnarray}
A_S 
&=& A_S^{\rm pert.}(z)+\frac{c}{\alpha b_0}\left(\frac{1}{\sqrt{3} A} \exp(-2 \alpha\, b_0 \, A_{\rm pert}(z))+ \exp(-2 \alpha\, b_0 \, A_{\rm pert}(z)) \right),
\end{eqnarray}
 where pert.  indicates   the case with $c=0$ discussed earlier. 
 Using Eqn. (38) leads to: 
 \begin{eqnarray}
 S_{NG}&=&  S_{NG}^{\rm pert}+ \frac{\,c}{\pi \alpha' \alpha b_0}\int_0^1\frac{d\eta}{\eta^2}\;(\frac{(a\eta)^{\alpha b_0}}{\sqrt{3}\log(a \eta)}+(a\eta)^{2\alpha b_0})\nonumber\\
 \delta S_{NG}^{\rm power}&=&   \frac{c}{\alpha' \alpha b_0\,(\alpha b_0-1)} a^{2\alpha b_0},
\end{eqnarray}
obtained after replacing $A_{\rm pert}\sim A_{AdS}$ in the evaluation.  Note that the power correction in this case did not stem from the contribution to $A_S$ by the dilaton, but rather from the correction order $c$ to the metric $A$ itself. This model gives a non-vanishing $G_2$ if $\alpha b_0=2$.

The non-perturbative model is now analyzed for $\alpha\geq 1$, where 
$\phi(u)=C e^{-\alpha u}$, $A(u)=u+\frac{C^2}{4\xi} e^{2\alpha u}$, and asymptotically $u=-\log z$.  Applying Eqn. (39) leads to:
\begin{equation}
S_{NG}=S_{NG}(AdS)+\frac{1}{\pi \alpha'}\int_0^1 \frac{d\eta}{\eta^2}\;(\frac{C}{\sqrt{3}} (a\eta)^\alpha+{\cal{O}} (a\eta)^{2\alpha}),
\end{equation}
where the term $\propto (a\eta)^\alpha$ stems from the contribution to $A_S$ by the dilaton. Clearly, if $\alpha=4$ the model gives a non-vanishing $G_2$, namely $G_2=\frac{4 C N }{\sqrt{3} \pi^3 \alpha' Z}$. For $\alpha=1$ it reproduces the additional logarithmic contribution in $z_0$ to the UV divergence in Eqn. (38). For $\alpha=2$ the model is similar to the one analyzed in 
   \cite{Andreev-Zakharov}. 
   In that case,  to obtain the $a^4$ power correction it is necessary to calculate to second order in the perturbation to the action, and therefore corrections to the solutions are to be calculated. As mentioned earlier, in QCD the power series in the area $s$ of the  Wilson loop   start at $s^2\sim a^2$; for $\alpha=2$ there is however a non-vanishing term order $a^2$  \cite{Andreev-Zakharov}.

\section{The trace anomaly test}

For the perturbative case the trace anomaly equation is clearly fulfilled.
Indeed the subtraction to the 5-dimensional action in  section V
was performed  in order to match, through Eqn. (1),
the vanishing of $G_{2}$ determined in the previous section for this model.
On the other hand, for the non-perturbative case, it is shown below that it is not possible
to match both sides of  Eqn. (1). 


\subsection{The trace anomaly equation for the non-perturbative case}

Equations (17) and (\ref{Tsub})  for
$\alpha\geq 1$ and $d=4$  lead asymptotically for $u_0\to \infty$ to:
\begin{eqnarray}
\frac{\beta(\lambda)}{\lambda} & =&-\frac{\alpha \phi}{(1+\frac{\alpha}{12}\phi^{2})}=-\frac{\alpha Ce^{-\alpha u_{0}}}{(1+\frac{\alpha C^{2}}{12}e^{-2\alpha u_{0}})}=-\frac{\alpha C z_{0}^\alpha}{(1+\frac{\alpha C^{2}}{12}z_{0}^{2\alpha})}\label{eq:beta-lambda}\\
T_{i}^{i}(sub.,NP) & = &-\frac{3}{8\pi G_{N}^{(5)}}(1-e^{\frac{C^2}{6} e^{-2\alpha u_0}}+\frac{\alpha C^2}{12}  e^{-2\alpha u_0})\nonumber \\
& =&-\frac{(\alpha+2)C^2}{32\pi G_{N}^{(5)}} z_0^{2\alpha} \label{tii},
\end{eqnarray}
where the asymptotic relation between domain wall and conformal coordinates
$z_{0}=e^{-u_{0}}$ has been employed. If the trace anomaly equation Eqn.
(\ref{trace-anomaly}) would be fullfilled, replacing  Eqns. (\ref{eq:beta-lambda})
and (\ref{tii}) into Eqn. (\ref{trace-anomaly}) would imply  for the gluon condensate to vanish asymptotically as:
\begin{equation}
G_2(z_0)=\frac{\alpha+2}{32 \pi^2  \alpha G_{N}^{(5)}}(C z_0^\alpha+C^2 z_0^{2\alpha}+\cdots).
\end{equation}

\subsection{Wilson loop calculation of $G_{2}$}

The computation of $G_{2}$ using the Wilson loop calculations of
the previous section involves a different choice of boundary conditions
than the one employed in this section. This is  because the Wilson
loop should be situated at a finite value of the coordinate orthogonal
to the boundary, corresponding to the finite value chosen in evaluating
the 5-dimensional action used to evaluate the trace of the energy-momentum
tensor. The boundary condition to be employed is,
\begin{equation}
z(a)=z_{0}\label{eq:bc}.
\end{equation}
For the pure AdS case, a solution of the area minimization equation
satisfying this boundary condition is given by,
$z(r)=\sqrt{a^{2}-r^{2}+z_{0}^{2}}$,
 which simply corresponds to a circle of radius $R=\sqrt{a^{2}+z_{0}^{2}}$
that is the radius required to match the boundary condition (\ref{eq:bc}).
For the non-perturbative model
the effect of the above mentioned change in boundary conditions
is well aproximated by replacing the radius $a$ by the effective
one corresponding to the AdS solution, i.e. $R=\sqrt{a^{2}+z_{0}^{2}}$.
Making that replacement in Eqn. (43) shows that the coefficient of $a^{\alpha}$
has a contribution, coming from the term proportional to $R^{\alpha}$
\footnote{Higher powers of $R$ also contribute to the coefficient of $a^{4}$,
giving contributions that vanish for $z_{0}\to 0$.
},  which does not vanish for $z_{0}=0$.  In particular, the simplest case where $\alpha=4$ gives a putative $G_2\neq 0$. This is however in contradiction with
the dependence in Eqn. (46),  which comes from assuming the
validity of (\ref{trace-anomaly}). Therefore,   the trace
anomaly equation is not fulfilled in this model, and this is so in general for $\alpha>1$.

\section{Conclusions and outlook}

In this work the validity of the trace anomaly equation has been studied in the holographic framework. This was  done by considering holographic evaluations of the VEV of the  trace of the energy-momentum tensor, the $\beta$-function and the gluon condensate $G_{2}$. The $\beta$-function is directly related to the definition of the particular model under consideration. The VEV of the trace of the energy momentum tensor was evaluated according to the holographic correspondence, by evaluating the $d+1$ dimensional classical action of the dilaton model on the corresponding classical solution.  The gluon condensate can be obtained in a YM theory from the VEV of the Wilson loop, which was here evaluated for the models studied by means of a NG action.

Two models were analyzed, which can be exactly solved and which have different qualitative characteristics. In the perturbative model, where $G_2=0$, consistency is fulfilled as the evaluation of the classical action can be appropriately subtracted to give a vanishing trace for the energy momentum tensor. If indeed $G_2=0$ in QCD, this may already be a somewhat  realistic model. On the other hand, the non-perturbative model shows an inconsistency for the trace anomaly equation. This is manifested by the fact that $G_2$ has different behavior in the scale $z_0$ in the two evaluations. Indeed, the evaluation of the action and  Eqn. (\ref{trace-anomaly}) give $G_2\propto e^{-\alpha u_0}=z_0^\alpha$, while from the Wilson loop evaluation $G_2$ is non-vanishing  in the limit $z_0\to 0$. This inconsistency seems   reasonable since the non-perturbative model fails to correctly describe the UV properties of QCD, being asymptotically AdS and not asymptotically free.

Various interesting conclusions can be drawn from these results. They indicate that, although a holographic model of the pure gauge QCD vacuum based on the AdS space is not feasible, they do not preclude a gravitational dual based on a dynamical 5-dimensional Einstein gravitational theory. They also show that QCD Ward identities, as for example the trace anomaly equation, strongly restrict the possibilities. It is  reasonable to expect that QCD symmetry restrictions can in principle lead to a more precise version of its putative gravitational dual. Such a dual should lead to a boundary theory having all the following properties: asymptotic freedom in the UV, confinement in the IR, (possibly) a non-vanishing gluon condensate, and consistency with the trace anomaly equation. As the examples considered have shown, it is not at all obvious how to obtain a consistent model with these properties. Work in this direction is in progress and will be reported in due course.

Among important fundamental non-perturbative effects in QCD, the existence of a non-vanishing gluon condensate was early on identified  \cite{SumRules}. It has important manifestations in hadron phenomenology \cite{SumRules,SumRulesReviews}, and there are indications of its non-vanishing from lattice QCD \cite{DiGiacomo-et-al,Rakow}. Due to its importance, its further understanding in the framework of holographic models of QCD is going to play a key role in the development of such models, as it has been shown in this work.

\section{Appendix}

This appendix presents an explicit calculation of the NG action in a case where the $A_S(z)$ has deviations from the AdS limit which are  integer powers of $z$, namely,
\begin{equation}
A_S(z)=-\log z+\sum_n  \alpha_n z^n.
\end{equation}
 The equation of motion Eqn. (32) is solved using an asymptotic series:
 \begin{equation}
 \omega(\eta)=\eta(1+\sum_n \sum_{\ell=0}^n C_{n\ell}\; \eta^n \log^\ell\eta),
 \end{equation}
where the coefficients $C_{n\ell}(\alpha_i,a)$ are obtained in a systematic fashion. 

The evaluation presented here can be applied to the non-perturbative model discussed in the text.
A straightforward but lengthy evaluation gives:
 \begin{eqnarray}
 S_{NG}&=&\frac{1}{2 \pi \alpha' }
\;   \left(\frac{a}{z_0}-\frac{8}{3} \,a \;\alpha_1 \log (z_0/a)+\frac{7}{18}\, a \;\alpha_1\right.\nonumber\\
   &+&a^2 \left(3.32435\;\alpha_1^2+\frac{11
}{3}   \;\alpha_2\right)\nonumber\\
   &+&a^3 \left(3.12395 \;\alpha_1^3+5.03896\; \alpha_2 \;\alpha_1+2\; \alpha_3\right)
\\
&+& a^4 \left(12.4174\; \alpha_1^4 
   +\left.19.5861\;
   \alpha_2 \;\alpha_1^2+2.09778\;\alpha_3
   \;\alpha_1+6.4849 \; \alpha_2^2+\frac{16 }{9}\;\alpha_4\right)  +{\cal{O}}(a^5)\right).\nonumber
   \end{eqnarray}
For instance, in a "soft wall" model where only $\alpha_2\neq 0$ one obtains:
\begin{equation}
 \omega^{\rm soft~wall}(\eta)=\eta(1+ \alpha_2 a^2 (-\frac{5}{3} \eta+\eta^2+\cdots)+\alpha_2^2 a^4 (-\frac{167}{27} \eta+\frac{125}{18}\eta^2+\cdots) +\cdots)
 \end{equation}
 and the resulting NG action becomes:
 \begin{equation}
 S_{NG}^{\rm soft~wall}=\frac{1}{2\pi \alpha'}(\frac{a}{z_0}+\frac{11}{3} \alpha_2 a^2+ \frac{134821}{20790} \alpha_2^2 a^4+\cdots)
 \end{equation}
 For the non-perturbative model with $\alpha=4$, one keeps only the term with $\alpha_4\neq 0$,
 and the NG action becomes:
  \begin{eqnarray}
 S_{NG}&=&\frac{1}{2 \pi \alpha' }
\;   \left(\frac{a}{z_0}+ \frac{16 }{9}\;\alpha_4 a^4    +{\cal{O}}(a^5)\right). 
   \end{eqnarray}

\section*{ACKNOWLEDGEMENTS}
The authors wish to  thank H. Casini, L. Da Rold, A. Rosabal and M. Schvellinger for  useful discussions and insights. This work was supported by DOE Contract No. DE-AC05-06OR23177 under which JSA operates the Thomas Jefferson National Accelerator Facility, by the National Science Foundation (USA) through grants PHY-0555559 and PHY-0855789 (JLG), and by CONICET (Argentina) PIP N¼ 11220090101018(RCT). JLG thanks the Instituto Balseiro and the Centro At\'omico Bariloche for hospitality during part of this project, the Programa Raices for sabbatical support through a Cesar Milstein Award. RCT thanks Jefferson Lab   for hospitality during part of this project.

\end{document}